# Inspection plan prediction for multi-repairable component systems using neural network

**Abstract ID: 789347**

Nooshin Yousefi[a], Stamatis Tsianikas[a], Jian Zhou[a] and David W. Coit[a],

[a] Department of Industrial and Systems Engineering, Rutgers University, Piscataway, NJ 08854-8065, USA

Email: no.yousefi@rutgers.edu

**Abstract**

Implementing an appropriate maintenance policy would help us to have a more reliable system and reduce the total costs. In this paper, a dynamic maintenance plan is proposed for repairable multi-component systems, where each component is subject to two competing failure processes of degradation and random shock. For systems with individually repairable components, it is not economical to replace the whole system if it fails. At any inspection time the failed components can be detected and replaced with a new one and the other components continue functioning; therefore, the initial age of each component at any inspection time is different from other components. Different initial ages have effect on the optimal time that the whole system should be inspected. The optimal inspection time should be calculated dynamically considering the initial age of all the components and their configuration within the system. In this paper, a neural network method is used to predict the next optimal inspection time for systems considering the initial age of components at the beginning of the inspection. System reliability and cost rate function are formulated and used to train the prediction model. The proposed maintenance plan is demonstrated by numerical examples.

## 1. Introduction

Most of the industrial and manufacturing systems have multiple components where each component degrades separately within the system. The components may fail due to multiple failure processes. The most common failure processes are degradation and random shock which can be considered as dependent competing failure processes. Failure due to continuous degradation is known as soft failure, and failure due to instantaneous stress caused by random shocks is called hard failure. It is assumed that each arriving shock may have some damages as additional abrupt change on the total degradation.

A repairable system is a system which after the failure of one or more components, can be restored to its satisfactory performance by replacing or repairing the system or components [1]. For many industrial systems with repairable components, the penalty cost due to downtime and loss of production is much higher than maintenance cost for system or each component individually. Finding an appropriate maintenance policy for each system would help the maintenance team to prevent the event of failure and, subsequently, the downtime cost [2]. However, the inspection interval is an important factor in each maintenance policy. Inspecting a system more frequently provides information about the status of the components and helps the maintenance team to prevent the failure; however, inspecting too often causes some unnecessary inspection and consequently increases the maintenance cost. Therefore, in this paper, a dynamic maintenance model is developed for a multi-component system with individually repairable components functioning for a very long time. In the proposed maintenance policy, the optimal next inspection time is found dynamically at the beginning of each inspection time for the whole system. It is also assumed that at each inspection time, the failed components are detected and replaced with a new one, while the other components continue to function.

## 2. Background

There have been several studies on reliability and maintenance of multi-component systems subject to degradation and random shocks. Yousefi et al [3] developed a reliability model for a system subject to mutually dependent competing failure processes. Shu and Flower [4] investigated the stochastic behavior of the reliability of repairable single systems. Lin et al [5] proposed a non-periodic condition-based maintenance policy for a deteriorating complex repairable system. Yi et al [6] studied reliability analysis of repairable systems with multiple fault modes. Rafiee et al [7] proposed a condition-based maintenance policy considering imperfect repair for a repairable deteriorating single system. Yousefi et al [8] proposed an on-condition maintenance model for a multi-component system where each component is subject to degradation and random shock while the components should be repaired as a group not individually. Zhu et al [9] developed a preventive replace-on-failure maintenance model for a *k*-out-of-*n*:F system subject to two competing failure processes. Fan et al [10] investigated a condition-based maintenance for a single unit repairable system subject to two statistically dependent failure modes which bidirectionally affecting each other. Le and Tan [11] studied a single unit deteriorating system whose condition is inspected periodically. Each degradation level can be represented by a state, making the system a multi-state system which is modeled by continuous-time Markov process. Most of the previous studies, consider a single system with only one repairable component. However, most of the systems consists of multiple repairable components where each of them can be degrades based on its own degradation rate. In this paper, we considered a deteriorating multi-component system with individually repairable components.

## 3. System reliability analysis

In this paper, it is assumed that each component is subject to two failure processes of hard failure and soft failure. Soft failure happens when the total degradation of component *i* is greater than its own predefined soft failure threshold ($H_i$). When any shock magnitude for component *i* is greater than a predefined hard threshold ($D_i$), component *i* experiences hard failure. Figure 1 shows these two failure processes for any component *i*.

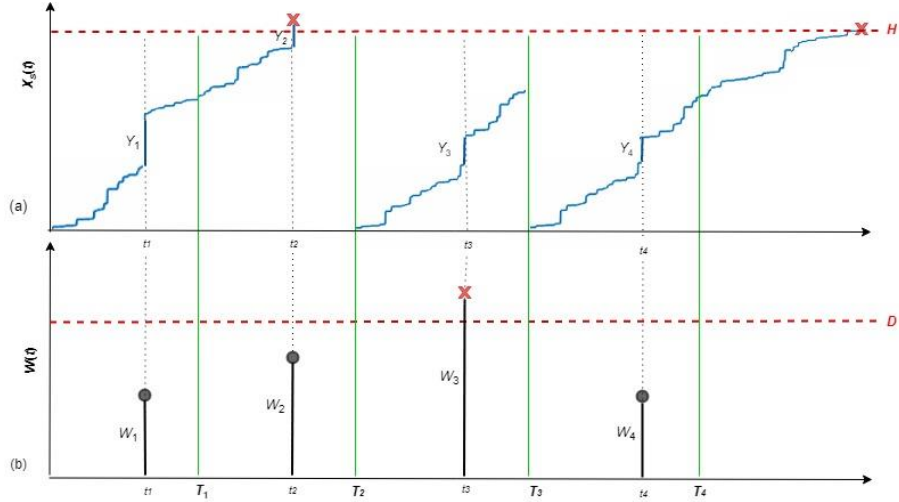

**Figure 1 (a)** soft failure process, **(b)** hard failure process[12]

In this study, it is assumed that the shock arrivals follow Poisson distribution and each shock has a magnitude following a normal distribution. $W_{ij} \sim Normal(\mu_{Wi}, \sigma^2_{Wi})$, where $W_{ij}$ is the $j^{th}$ shock magnitude of component *i*. Therefore, the probability of having no hard failure for component *i* can be calculated as follow:

$$P_{NH,i} = P(W_{ij} < D_i) = F_{W_i}(D_i) = \Phi(\frac{D_i - \mu_{W_i}}{\sigma_{W_i}}) \quad (1)$$

The probability of having *m* shocks by time *t* is given as follow, where *λ* is the parameter of Poisson distribution as the shock arrival rate.

$$P(N(t) = m) = \frac{(\lambda t)^m e^{-\lambda t}}{m!} \quad (2)$$

To calculate the probability of having no soft failure, we need to model the degradation process of each component. In this paper, gamma process is used, which is a suitable stochastic process for monotonic increasing degradation path; So, $X(t)-X(s) \sim gamma(\alpha(t)-\alpha(s), \beta)$, where $X(t)$ is the degradation level at time *t*, $\alpha(t)=\alpha \times t$ is gamma shape parameter which is linear in *t*, and *β* is scale parameter.

$$g(x; \alpha_i(t-s), \beta_i) = \frac{\beta_i^{\alpha_i(t-s)} x^{\alpha_i(t-s)-1} \exp(-\beta_i x)}{\Gamma(\alpha_i(t-s))} \quad (3)$$

Since each incoming shock has a damage on the degradation path of all the components. we defined a random variable for damage shock which follows normal distribution $Y_{ij} \sim Normal(\mu_{Yi}, \sigma^2_{Yi})$, where $Y_{ij}$ is the $j^{th}$ shock damage on the $i^{th}$ component. Therefore, the probability that component $i$ has not experienced soft failure by time $t$ is as follow, where $S_i(t)$ is the summation of damages on component $i$ by time $t$. The total degradation is the pure degradation and all the damages from shock, $X_{S_i}(t) = X_i(t) + S_i(t)$.

$$P_{NS}(t) = P(X_{S_i}(t) < H_i) = \sum_{m=0}^{\infty} P(X_i(t) + S_i(t) < H_i) \times \frac{(\lambda_0 t)^m e^{-\lambda_0 t}}{m!} \quad (4)$$

In this paper, it is assumed that each failed component is detected just by inspection. Since we replace the failed component instead of the whole system, the age of components at each inspection time is different from other. Yousefi and Coit [13] studied a multi-component system with individually repairable components. This paper is an extension of [13]. For the dynamic maintenance policy, the inspection interval should be found dynamically based on the initial age of all the components. For a system with multi-components that degrading differently, a preventive maintenance model should be found considering the age of all the components at the beginning of each interval.

Figure 2 shows a dynamic maintenance model for a system with three components. At the beginning of each inspection time the age of all the components are detected and the new inspection time is found dynamically. Random values $u_i$ is assumed as the initial age of each component $i$ at the beginning of interval. In [13], the cost rate function is defined for maintenance of the system and by solving the optimization problem at each inspection time the next inspection duration is found.

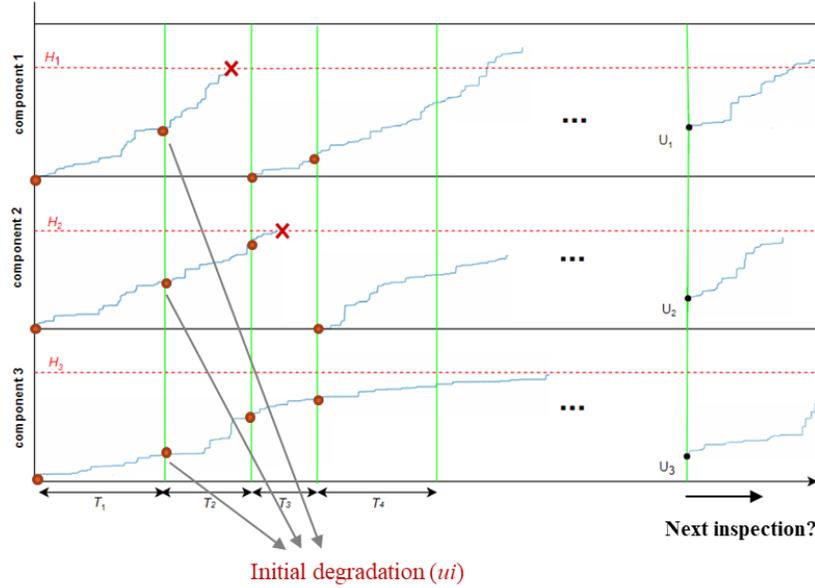

**Figure 2** Dynamic inspection planning considering random initial age for each component [13]

Using Equation (4), considering the of initial age of component I, the probability of having no failure by time $t$ can be calculated as follow where $u_i$ is the initial degradation level at the beginning of the inspection interval:

$$R_i(t; u_i) = \sum_{m=0}^{\infty} \left[ P(W_{ij} < D_i)^m P\left(X_i(t) + S_i(t) + u_i < H_i\right) | N(t) = m \right] \frac{(\lambda_0 t)^m e^{(-\lambda_0 t)}}{m!} \quad (5)$$

$$= \sum_{m=0}^{\infty} \left[ P(W_{ij} < D_i)^m \times \int_0^{H_i} P\left(X_i(t) + y + u_i < H_i\right) f_{Y_i}^{<m>}(y) dy \right] \frac{(\lambda_0 t)^m e^{(-\lambda_0 t)}}{m!}$$

For each system configuration the system reliability should be calculated considering the failure behavior of system. For a series system, failure of any component causes the system failure while in a parallel configuration, all the components should fail to make the whole system fails. Therefore, the system reliability by time $t$, for a series system ($R_s(t)$) and for a parallel system ($R_p(t)$) considering the initial age of all the components can be calculated as follow:

$$R_S(t;\boldsymbol{u}) = \sum_{m=0}^{\infty} \prod_{i=1}^{n} \left[ P(W_{ij} < D_i)^m \times \int_0^{H_i} P\big(X_i(t) + y + u_i < H_i\big) f_{Y_i}^{<m>}(y) dy \right] \frac{(\lambda_0 t)^m e^{(-\lambda_0 t)}}{m!} \quad (6)$$

$$R_P(t;\boldsymbol{u}) = 1 - \sum_{m=0}^{\infty} \prod_{i=1}^{n} \left[ 1 - \left( P(W_{ij} < D_i)^m \times \int_0^{H_i} P\big(X_i(t) + y + u_i < H_i\big) f_{Y_i}^{<m>}(y) dy \right) \right] \times \frac{(\lambda_0 t)^m e^{(-\lambda_0 t)}}{m!} \quad (7)$$

Yousefi and Coit [13] formulated a cost rate function for maintenance of a multi-component system and by solving the optimization problem with objective function of cost rate, the optimal inspection time as a decision variable is found. However, solving the optimization problem at each inspection time is not time and computational efficient. In the proposed method, a neural network method is used to predict the inspection time of the system at the beginning of each inspection time. The neural network model can be trained using the optimal inspection times calculated by [13]. Neural network is one of the most widely used machine learning methods. The neural network models are developed to mimic the decision-making process of human beings and do not require users to predefine a mathematical equations [14]. Therefore, in this paper, by using a neural network model, the maintenance team can find the best optimal inspection time at each decision epochs without solving an optimization problem. Figure 3 shows the neural network model for the proposed maintenance model. The input of the network should be the initial age of all the components in the system and the information about the shock process and degradation process of all the components.

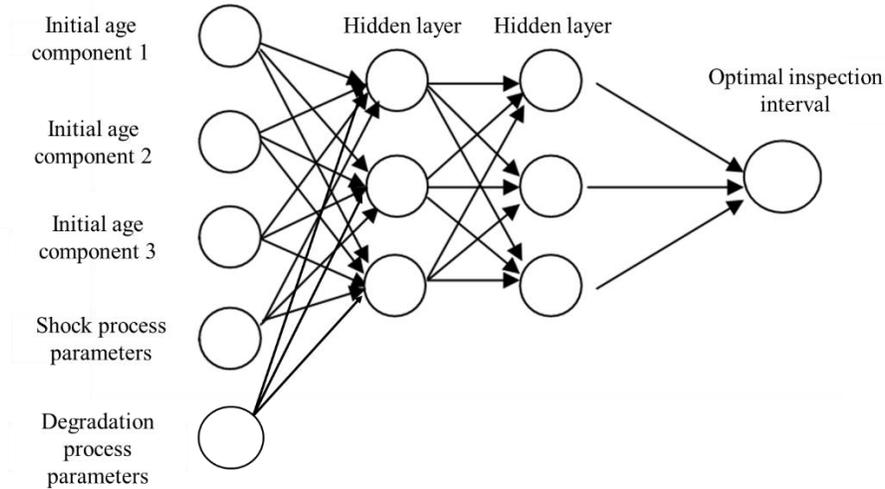

**Figure 3** Neural network model for maintenance problem

The neural network model has different hidden layers that each has different nodes. A node is a place where computation happens. At each node, the inputs of the model are combined with some weights, and then sum of all the input-weights products is passed through an activation function to transitions to the next layer. At the end of this process, the last hidden layer is linked to the output layer to provides the output. One of the main strengths of machine learning algorithms is their ability to learn and improve every time in predicting an output.

$$output = f(z_1 \times \theta_1 + z_2 \times \theta_2 + ... + z_p \times \theta_p + b) \quad (8)$$

Equation (8) shows the calculation of output of each layer when there are $p$ inputs into the model. $b$ is the bias and $f(.)$ is the activation function. The activation function is used to convert the inputs into a predictable form of output. In this study Sigmoid activation function is used for the proposed maintenance neural network. To evaluate the output of a neural network is a cost function should be calculated to measure how the prediction is good. One of the commonly used cost functions is mean squared error which is calculated as follow:

$$MSE = \frac{1}{n} \sum_{j=1}^{n} (y_{true} - y_{pred})^2 \quad (9)$$

where $n$ is the number of samples, and $y_{true}$ is the actual value of which is the inspection interval and $y_{pred}$ is the predicted value or the predicted next inspection interval in our problem. By calculating the cost function, the weights and biases of the network can be optimized. Stochastic gradian decent (SGD) is one of the optimization algorithms to minimize the lost function by changing the weights and biases. The update equation for SGD is:

$$\theta_{k,j+1} \longleftarrow \theta_{k,j} - \eta \frac{\partial L}{\partial \theta_{k,j}} \qquad (10)$$

$L$ is the lost function, and $\eta$ is the learning rate that controls the speed of training. $\theta_{k,j}$ is the weight $k$ for time step $j$. By using SGD and finding the optimal weights and biases on the training sample data, a neural network is training and can be used for the future predictions. In this paper, a neural network model is trained on different combinations of initial degradation levels of all the components at the beginning of the inspection interval and parameters of degradation and shock processes, and the next inspection time can be predicted as the output of the model. By using the trained neural network model, there is no need to solve an optimization problem to find the optimal next inspection time, and it can be computationally efficient.

## 4. Numerical results

To show the preference of the proposed method, a conceptual example is considered to demonstrate the proposed dynamic maintenance model. The numerical example is a series system with three components. Table 1 shows the parameter assumptions for this example. Based on model presented in [13] the optimal inspection interval for these systems are found by solving an optimization problem for the maintenance cost rate at the beginning of each inspection time for 60 different scenarios. To show the preference of proposed model, a neural network model is trained on the result of optimization problem. By splitting the dataset into two groups of testing and training, the model is trained on 70% of the data and tested on 30% of them.

**Table 1** Parameter values for numerical examples

| Parameter | description | component 1 | component 2 | component 3 |
|---|---|---|---|---|
| $H$ | The soft failure threshold | 20 mm | 30 mm | 35 mm |
| $D_i$ | The hard failure threshold | 7 | 5 | 6 |
| $\alpha_i$ | The shape parameter for gamma process | 3 | 2 | 1 |
| $\beta_i$ | The scale parameter for gamma process | 1 | 0.6 | 0.3 |
| $\lambda$ | The initial intensity of random shocks | | $2.5 \times 10^{-3}$ | |
| $Y_{ij}$ | Shock damage | $Y_{ij} \sim N(2, 0.5^2)$ | $Y_{ij} \sim N(2.5, 0.2^2)$ | $Y_{ij} \sim N(3, 0.1^2)$ |
| $W_{ij}$ | The shock magnitude | $W_{ij} \sim N(1.5, 0.4^2)$ | $W_{ij} \sim N(2, 0.3^2)$ | $W_{ij} \sim N(1.2_i, 0.15^2)$ |

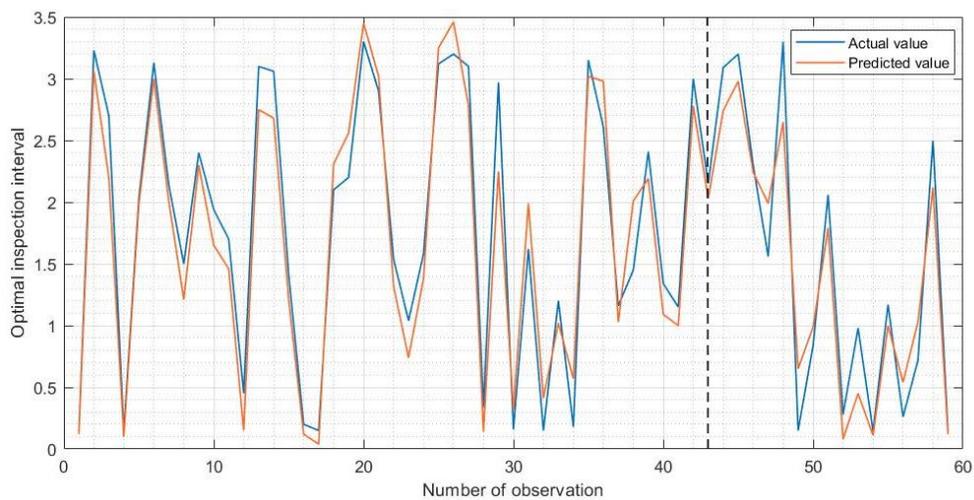

**Figure 4** optimal inspection interval versus the predicted using neural network

Figure 4 shows the predicted inspection interval using a neural network versus the optimal inspection interval using the model in [13]. The R-squared value for this model is 0.93 which shows the proposed method can provide the similar result as an optimization problem. Moreover, the time for finding the optimal value of the next inspection time by solving the optimization problem is 28 minutes, while the proposed neural network provides the prediction in less than 1 minute. Using the neural network to find the next inspection interval would help the maintenance team to save time and money due to computational saving. In figure 4, the vertical dashed line separated the train and test observations. The left side of the dash line shows the training data used to train the neural network, and the right side shows the test observations.

## 5. Conclusion

In this paper, a new dynamic maintenance model is developed for a system with multiple components, where each component can be repaired individually within the system. It is assumed that each component can fail due to two failure modes, which are competing and dependent. Degradation process and random shock process are the main competing failure processes that are considered in this model. Each component is degrading separately and subject to two competing failure processes, and at each inspection interval, the failed component can be replaced with the new ones, while the other components can continue functioning. The new dynamic model is an extension of the previous models by combing the traditional optimization problem with a neural network model. Using a Neural network would lead us to find the next inspection interval without solving a maintenance optimization problem, which is more computationally efficient. The numerical results show that the predicted next inspection interval is very similar to optimal values, which can be obtained by solving the optimization problem while the time to find the next inspection time using the proposed neural network is less than solving the traditional optimization problem.